\documentclass[conference]{IEEEtran}
\IEEEoverridecommandlockouts
\usepackage{cite}
\usepackage{amsmath,amssymb,amsfonts}
\usepackage{algorithmic}
\usepackage{graphicx}
\usepackage{textcomp}
\usepackage{xcolor}
\usepackage[T1]{fontenc} 
\def\BibTeX{{\rm B\kern-.05em{\sc i\kern-.025em b}\kern-.08em
    T\kern-.1667em\lower.7ex\hbox{E}\kern-.125emX}}

\usepackage{fancyhdr}

\begin{document}

\title{Evolution of Technologies and Multivalued Circuits}

\author{\IEEEauthorblockN{ Daniel Etiemble}
\IEEEauthorblockA{\textit{Computer Science Laboratory (LRI)} \\
\textit{Paris Sud University}\\
Orsay, France \\
de@lri.fr}

}

\maketitle

\begin{abstract}
For more than 45 years, many multi-valued circuits have been presented. With  very rare exceptions, they have been unsuccessful for fundamental reasons that can be explained. Each time a new circuit technology is presented, a lot of new MVL circuits are proposed. Can new circuit technologies overcome the fundamental disadvantages of MVL circuits? The evolution of IC technologies in the last decades unfortunately increases the disadvantage of MVL circuits versus binary ones. For non conventional technologies, only quantum devices look promising, even if implementation is challenging and applications are restricted to a small niche.
\end{abstract}


\section{Introduction}
M-valued circuits are just between binary circuits (M=2) and analog circuits (M=$\infty$). In the last decades,
many analog implementations have been replaced by binary implementations (radio, TV, photography, cinema, etc.).
M-valued circuits are closer to binary circuits than analog ones. They could have taken advantage from the shift towards
digital implementations.

A recent survey of ``Contempory Aspects of Multiple-Valued Logic and Its application to Microelectronics Circuits'' can be found in \cite{Gaudet}. Many M-valued circuits have been proposed and some have been fabricated and tested for the last 45 years. Several M-valued
 Flash and DRAM memories have been integrated in the 90s \cite{Gulak}. Even with
promising performance, they have not been able to compete with binary ones on the long term, flash memories being the exception. 

Obviously, M-valued circuits could be interesting only if they provide significant advantages versus binary ones. It means that every proposed new M-valued circuits should be compared with the corresponding binary ones. That is this approach that we used in \cite{b1}  which also contain references of M-valued circuits proposed in the 80s and 90s. M-valued circuits must also use a technology that is compatible with the standards of up-to-date foundries.

In this paper, we first summarize the basic reasons for which M-valued circuits are generally more complex and less efficient than the binary ones. Then we examine the evolutions of Integrated Circuits Technologies in the last 45 years to determine if the main trends can overcome the intrinsic disadvantage of M-valued circuits. Then we discuss the case of quantum circuits for which a qubit holds up to two bits with super dense coding.

\section{Binary and M-valued circuits in classical computer technologies }
\begin{figure}[htbp]
\centerline{\includegraphics  [width = 8 cm]{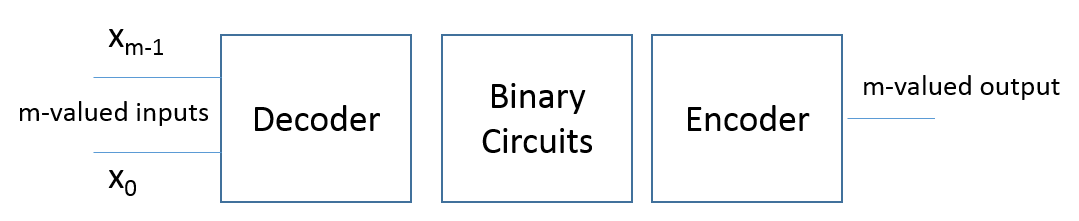}}
\caption{General scheme of M-valued circuits.}
\label{fig}
\end{figure}

A detailed analysis of the comparison has already been presented in the paper ``Why M-valued circuits are restricted to a small niche'' \cite{b2} published in 2003. We just summarize the main points and show that the situation has not changed since 2003.
While bits have two states (0 and 1), M-valued  circuits have m different states, generally from 0 to m-1. The different states can be several levels of voltages, or currents or level of charges. The key point is that these levels are totally ordered. When considering m=4=2$^2$, it means that three threshold detectors are needed to detect each value while only one is needed with binary circuits. When m=2$^k$, which is the case for a simple interface with binary circuits, there are k-1 detectors versus $\log_2 k$ due to the lattice property of Boolean algebra. The general scheme of M-valued circuits is presented in Fig. \ref{fig}.  Current-mode M-valued circuits can use analog sum or difference of  currents, but need current mirrors to duplicate output current values. 

Interconnection issues have been quoted for binary circuits: they are probably the most important argument for promoting M-valued circuits. Let's consider again Fig \ref{fig}. There are two possibilities:
\begin{itemize}
\item Each circuit gate or building block is implemented as an M-valued circuit. It means that each one obeys to the diagram presented in Fig. \ref{fig}: encoder, binary circuits and decoder. Then, it is easy to demonstrate that the M-valued circuits use more transistors and more internal interconnects than the corresponding binary circuits. A detailed comparison will be presented in section IV-B for Fig \ref{Navi}.
\item M-valued interconnects are only used to reduce the number of interconnects between building blocks or even between different chips. In Fig. \ref{fig}, all the computations are done with binary circuits. Reducing the number of interconnects with multiple levels  is used in amplitude modulation: for instance, PAM-4 coding, that uses 4 levels to code 2 bits is adopted for high-speed data transmission (IEEE802.3bs). However, this approach is not used by computer designers.  Different types of high-speed serial links are used such as SuperSpeed USB, PCI, XAUI, Infiniband, RapidIO and SATA. NVidia and IBM use NVLink, Intel uses QuickPath. These serial links use differential signaling, doubling the number of wires compared to single-end signaling.
\end{itemize}

Another argument used for M-valued circuits is that radix R = 3 would be more economical than R = 2 because the ``optimal'' radix is R = e = 2.718. The demonstration can be found in \cite{Hurst}:
The number of digits necessary to express a range of N is given by N=$R^d$, where R is the radix and d the number of digits, rounded to the next highest value. It is assumed that the complexity C of the system hardware is proportional to the digit capacity $ R \times d$ where k is a constant.

\begin{equation}
C = k  (R \times d) = k(R \times  \frac {log N}{log R})
\label{eq}
\end{equation}

Differentiating with respect to R shows that R = e for a minimum cost C. The problem with \eqref{eq} is that the cost is only considered as proportional to the digit capacity $R \times d$, implicitly assuming that the hardware cost is the same for any value of R.  Let assume that the hardware complexity is proportional to R. It looks more realistic as the number of levels is proportional to R and the number of threshold detectors  proportional to $R -1$.   In that case, with a different $k_2$ constant, the new equation is:

\begin{equation}
C = k_2 R (R \times d) = k(R^2 \times  \frac {log N}{log R})
\label{eq0}
\end{equation}

Now, differentiating with respect to R shows that $R= \exp {(0.5)} = 1.645$, to be rounded to the next integer R = 2. This suggests that the binary system is the most efficient. This result has an advantage: it corresponds to what has been observed since more than  five decades!

It does not mean that ternary or M-valued circuits must never be used: they can and must be used when they implement some functions having 3 or m different states. A good example is the ternary content-addressable memories (CAM) which cells store three different states: 0, 1 and X (don't care). It allows to search for words having unknown and non-significant bits. However, implementing three states in CAM cells faces the problems of M-valued circuits presented in the introduction.

While there are huge advantages for binary circuits, it is still interesting to evaluate how the evolution of technologies have influenced and will influence the ``competition'' between M-valued and binary circuits.

\section{Evolution of typical semi-conductor Technologies}
While Interconnection issues have been quoted by circuits designers as a reason for proposing M-valued circuits, it turns out that the power dissipation has been the main factor driving the evolution of technologies and circuitries, even before the ``heat wall'' has been coined \cite{b3}.

For any technology, power dissipation is decomposed into static and dynamic parts (equation \ref{eq1}).
\begin{equation}
Pd =Pd_{static} + Pd_{dynamic} \label {eq1}
\end{equation}

For CMOS technologies, power dissipation is given by equation \ref {eq2}. In this equation, V$_{dd}$ is the supply voltage, $\alpha$ is the fraction of the capacitances that are switching (activity factor),  $\sum{ C_i }$ is the sum of capacitances and F is the clock frequency. In this paper, V$_{dd}$ is the main factor in the comparison between binary and multivalued circuits.

\begin{equation}
Pd =V_{dd} \times I_{leakage} + \alpha \times \sum{ C_i  \times V_{dd}^2 \times  F} \label{eq2}
\end{equation}

\subsection{Bipolar technologies}
In the 70s and 80s, several M-valued bipolar circuits have been presented corresponding to the currently used bipolar circuits of this time: 
\begin{itemize}
\item Voltage mode ternary TTL circuits have been presented in \cite{b4},
\item Voltage mode 4-valued ECL encoder and decoder circuits have been presented in \cite{b5},
\item Current mode multivalued I2L circuits have been presented in \cite{b6}.
\end{itemize}

\begin{figure}[htbp]
\centerline{\includegraphics  [width = 8 cm]{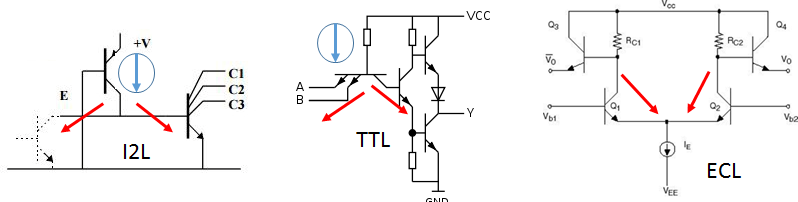}}
\caption{Bipolar circuitries}
\label{fig2}
\end{figure}

The binary circuitries of I2L, TTL and ECL logic families are presented in Figure \ref{fig2}. While ECL uses an actual current generator, I2L and TTL use a pseudo current generator. In both cases, the corresponding current flows towards input or output for I2L and TTL, and towards left or right sides of the differential pair for ECL. A significant current contributes to the static power dissipation: it is not needed to consider dynamic power dissipation to understand why bipolar digital circuits have disappeared as soon as MOS technology has  become mature and CMOS circuitry, that had no static power dissipation at that period, was able to replace pMOS and nMOS circuitries.

Significant static power dissipation has  excluded the pMOS and NMOS versions for which a current flows from power supply to  ground when the output transistor is on. It also excludes any binary or M-valued CMOS version that would be based on a differential pair as in ECL circuits. It also excludes current mode circuits that are based on different levels of currents.

\subsection{CMOS circuits}
\begin{figure}[htbp]
\centerline{\includegraphics  [width = 8 cm]{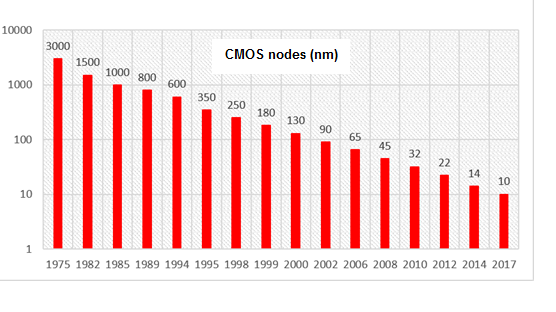}}
\caption{CMOS technological nodes}
\label{figurenode}
\end{figure}

\begin{figure}[htbp]
\centerline{\includegraphics  [width = 8 cm]{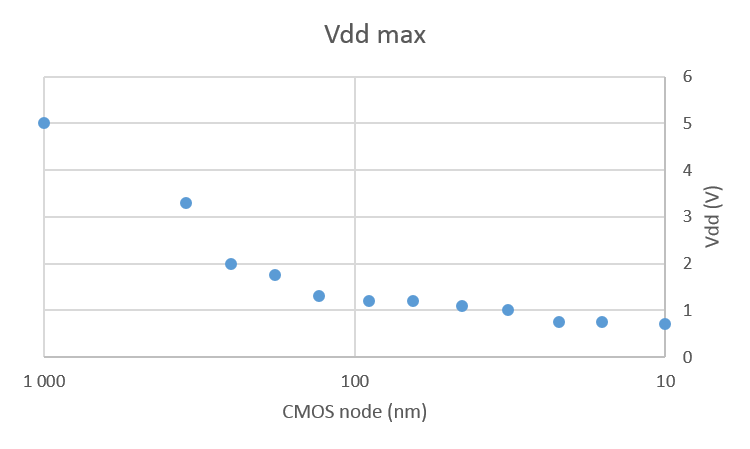}}
\caption{V$_{dd}$ scaling since 1000-nm node}
\label{Vddmax}
\end{figure}

The doubling of the number of transistors every N months (12, then 18, then 24) according to Moore's law has been realized by regularly launching a new generation of CMOS technologies, called a technological node. The different successive nodes are shown in Figure \ref{figurenode}. It is out of the scope of this paper to discuss all the consequences of the scaling of successive nodes.
For M-valued circuits, two consequences should be outlined:
\begin{itemize}
\item Power supply values have been reduced, from 12 V for the first pMOS circuits down to 5V and down to a value in the 0.8 to 1V range in every technology used since 2006 (65 nm node).  Fig. \ref{Vddmax} presents the scaling of  V$_{dd}$ since the 1000 nm node. In 2000 with 130 nm node, V$_{dd}$ value was already as low as 1.3 V. This scaling of V$_{dd}$ means that the voltage swing available for implemented M-valued circuits is reduced to the minimal value used by binary circuits.
\item For CMOS circuits, power dissipation is given by equation \ref{eq2}. Since the late 90s,  CMOS static power dissipation is no longer negligible: it is proportional to V$_{dd}$. Dynamic power dissipation is proportional to V$_{dd}^2$. Obviously, reducing power dissipation means reducing   V$_{dd}$ and using the smallest  V$_{dd}$ value compatible with a correct behavior of transistors. It is another reason for the scaling down of V$_{dd}$. 
\end{itemize}

\subsubsection{CMOS voltage mode M-valued circuits}
As the smallest  V$_{dd}$ value compatible with a correct behavior of transistors must be used to reduce power dissipation, it means that the different values of M-valued circuits must be included in the  V$_{dd}$ range: 0, V$_{dd}/2$ and V$_{dd}$ for ternary circuits or 0, V$_{dd}/3$,  2 V$_{dd}/3$ and V$_{dd}$ for quaternary circuits. This is more and more difficult with reduced  V$_{dd}$  values.

\subsubsection{CMOS M-valued memories}
A review of multiple-valued memory technology has been published in the late 90s \cite{Gulak}. M-valued ROM, Flash, DRAM and CAM have been implemented and tested. It is probably the most successful area for M-valued circuits. 

For M-valued ROMs, two techniques can be used:
\begin{itemize}
\item  The first technique stores one out of four states within a single cell, keeping cell size unchanged. Each cell consists of a MOS transistor having one out of four impedance values $Z_0<Z_1<Z_2<Z_3$, which correspond to different transistor channel lengths, programmed at the diffusion or polysilicon level. Reading a cell is done by comparing the cell impedance with reference transistor impedance $Z_{0.5}, Z_{1.5}$ and $Z_{2.5}$. The overall cell area is thus divided by two compared to binary ROMs, with an overhead for the comparator and decoder circuits. Details on different old commercial circuits can be found in \cite{Gulak}..
\item The second technique also store one of four states in a single cell, but with a transistor having one of four different threshold voltages. The threshold detection of a cell is realized by linearly ramping the input of the transistor. When the input reaches the threshold level of this transistor, it turns on. This approach can be used when the access time is not the objective.
\end{itemize}

M-valued DRAM have also been presented. $2^N=M$ different possible charges are stored in the capacitance $ C_s$ of a DRAM transistor cell. In the write-mode operation, a descending $2^N$-level staircase is applied is applied to the word line, i.e. to the selected transistor. For the $ i^{th}$  level input, the data line voltage is changed from low to high when the pulse level is i. In the read-mode operation, an ascendant staircase pulse is applied to the word line. The same pulse is transferred to a dummy cell, which transistor has a $C_s/_2$ capacitance for signal comparison: it is obvious that cycle time to read or write is long, as it depends on the number of levels of the staircase pulse. If density is multiplied by N, a part of the read or write cycle time is multiplied by $2^N$ while the other part corresponds to the charge transfer preamplifier and the sense amplifier timing characteristics.

M-valued SRAMs and DRAMs  have been fabricated and tested by industrial companies in the 80's and the 90's, such as Hitachi \cite{Aoki}, NEC \cite{Sugi} and \cite{Murotani}, etc. They are detailed in \cite{Gulak}. During the following decades, there was no longer such presentations by industrial companies. The main reason is that the  V$_{dd}$ level that is used by the successive nodes is too small to use the previously used techniques with conventional technologies.

M-Valued flash memories also use similar techniques and were presented in the 90s \cite{Bauer,Jung}. They are largely used.  4-valued (MLC) flash memories store two bits per cell. 8-valued (TLC) memories store 3 bits per cell. In 2018,  ADATA, Intel, Micron, and Samsung have launched some SSD products using QLD NAND-memory with 4 bits per cell. While binary flash memories have the advantage of faster write speeds, lower power consumption and higher cell endurance, M-valued flash memories provide higher data density and lower costs

,

\section{What about new technologies?}
\subsection{Foreword}
For many years, the end of Moore's law has been announced. In 2003, G. Moore had anticipated the ``end'' of the exponential growth of transistors by quoting: ``No exponential is forever: But "forever" can be delayed". While the fundamental limits are approaching, new technological nodes  have already been announced. In 2017, Intel, Samsung and TSMC provided a 10-nm technology. In 2019, TSMC provides a 7-nm technology. 5-nm and 3-nm nodes have already been announced. In June 2017, IBM presented the first chip with a 5-nm node. TSMC announced the 5-nm delivering in 2020 and a new fab to be built for a 3-nm delivering in 2022. The limits are approaching, but are still some years ahead!

It does not mean that new technologies are useless: it means that the competition is severe. Any new technology has to prove its efficiency in term of performance, cost, reliability in a large spectrum of applications against the modern FinFET or SOI CMOS. Except if the technology exhibits actual M-valued behavior, such as quantum devices, it is doubtful that the M-valued circuits will be more efficient than the binary ones as they are faced to the issues that have been previously called to mind in this paper.

Too often, as soon as a new technology gains some popularity, some researchers just present a new version of old M-valued circuits with the new technology. Sometimes, the result is awful. This  technology can exhibit some drawbacks that prevent it to become a mature technology. The corresponding circuitry can exhibit a major drawback, such as a static power dissipation like the nMOS circuitry.

\subsection{CNTFET technology}
A carbon nanotube field-effect transistor (CNTFET) refers to a field-effect transistor that uses a single carbon nanotube or an array of carbon nanotubes as the channel material instead of bulk silicon in the traditional MOSFET. The MOSFET-like CNTFETs having p and n types look the most promising ones. The technology has advantages and drawbacks:
\begin{itemize}
\item CNTFET have variable threshold voltages (according to the inverse function of the diameter), which is a great advantage to implement M-valued circuits. Among other advantages, high electron mobility, high current density, high tranductance can be quoted.
\item Lifetime issues, reliability issues, difficulties in mass production and production costs are quoted as disadvantages.
\end{itemize}
CNTFET is a technology far from being able to compete with the most advanced CMOS technologies:
\begin{itemize}
\item In 2012, IBM announced a breakthrough in nanotube computer chip fabrication. The IBM researchers announced: ``Carbon nanotubes have the potential in the development of high-speed and power-efficient logic applications. However, for such technologies to be viable, a high density of semiconducting nanotubes must be placed at precise locations on a substrate'' and ``This new placement technique is readily implemented, involving common chemicals and processes, and provides a platform for future CNTFET experimental studies''. However, no further information has been provided by IC manufacturers since this annoncement.
\item In 2013, the first carbon nanotube computer has been announced \cite{Shulaker}. It is a significant advance for this technology. However, this 178 CNTFETs ``one-instruction-set computer" only runs at 1 KHz. The first commercial microprocessor (Intel 4004) had 2300 transistors and run at 780 KHz in 1971. This illustrates the difference for a first computer and the gap to compete with to-day microprocessors. 
\end{itemize}
For implementing MVL circuits, the MOSFET-like CNTFETs have a circuitry similar to CMOS circuitry, but with an easier way to implement different threshold voltages. Different M-valued CNTFET circuits have been proposed. The most significant is probably \cite{Navi} published in 2015. A similar paper by the same authors has been published in 2016. Both papers compare the ternary or 4-valued proposed circuits with previously proposed ternary or 4-valued CNTFET circuits. However, there is no comparison between these M-valued circuits and the corresponding binary circuits. 

\begin{figure}[htbp]
\centerline{\includegraphics  [width =8 cm]{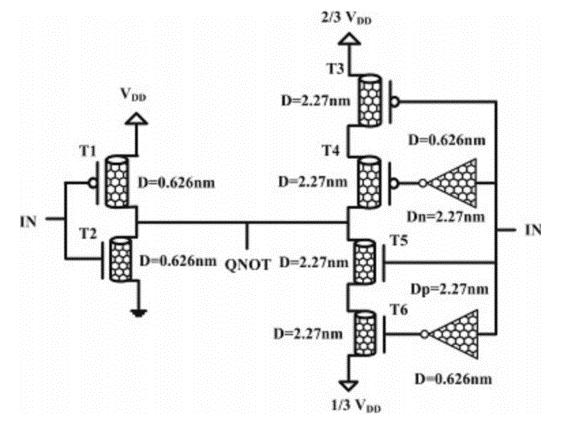}}
\caption{4-valued inverter proposed in \cite{Navi}}
\label{Navi}
\end{figure}

The 4-valued inverter circuit presented in Fig. \ref{Navi} allows this comparison. Let's assume that a 4-valued inverter would correspond to two binary inverters. A binary inverter has two transistors. The 4-valued inverter has ten transistors. There is a 10/4 = 2.5 advantage for the binary inverter. Without considering input and output  connections, a binary inverter has four connections. A quaternary inverter has twelve connections for the three binary inverters, five connections between 2/3 Vdd and 1/3 Vdd,  six connections for the right IN part and one connection between left and right parts (QNOT), for an overall 24 connections. Assuming again that two binary inverters correspond to one 4-valued inverter, there is a 24/4 = 6 advantage for the binary inverters. It is somewhat difficult to argue that M-valued circuits reduce the number of interconnections. In this example, the number of external interconnections is divided by two for the 4-valued inverter while the number of internal interconnections is multiplied by six. Similar results would be derive from comparing quaternary NAND or NOR. 

The M-valued CNTFET circuits do not provide any advantage versus CNTFET binary circuits: they have the same intrinsic disadvantage that was mentioned in section II.

\subsection {Single-Electron Transistor}
\begin{figure}[htbp]
\centerline{\includegraphics  [width =8 cm]{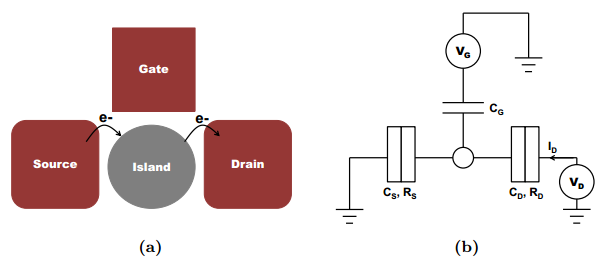}}
\caption{(a)Schematic of SET ; (b) Schematic of equivalent circuit}
\label{SET}
\end{figure}

Fig. \ref{SET} presents the schematics of a Single-Electron Transistor (SET). This device can be used in M-valued circuits or M-valued memories as it has several thresholds in the non-monotonic, oscillatory $I_D$-$V_G$ characteristics.

A lot of papers have been presented in the 90s and far less in the  last two decades. SET devices had to solve two types of problems:
\begin{itemize}
\item Being able to operate at normal temperature, as SET devices don't exhibit such performance advantage to justify to operate a very low temperature like quantum devices that we will consider in the next section.
\item Being combined with CMOS technology to overcome the intrinsic limits of SET-only technology.
\end{itemize}
Recent papers such as \cite{CEA} show that the two issues can be solved, at least in the French CEA Technological research center for industry.

\begin{figure}[htbp]
\centerline{\includegraphics  [width =8 cm]{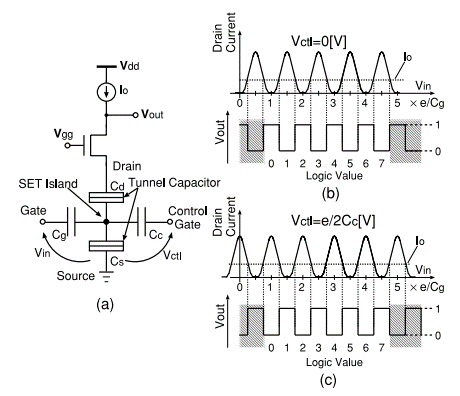}}
\caption{SET periodical literal circuit (a)Schematic,  (b) and (c) transfer characteristics for two $V_{ctl}$ values}
\label{SETP}
\end{figure}

The multithreshold transfer characteristics can thus be used to implement M-valued circuits. Fig. \ref{SETP} shows the SET periodical literal circuit  from which different M-valued gates can be derived \cite{Degawa}. Actual comparisons between M-valued performance and the corresponding binary ones are still to be done.

While SET technology has improved in the recent period, it is still far from being a serious competitor for classical FinFET or SOI CMOS technologies. 

\section{Other technologies}
This paper does not pretend to consider all the technologies likely to allow the realization of M-valued circuits. For instance, a recent paper shows the potential of Resistive random access memory (ReRAM) \cite{Batt} for implementing ternary logic. These technologies could be considered in a further paper.

\section{Quantum Computers}
As previously mentioned, one of the main issues of M-valued circuits with integrated circuits technologies is that having more than two states needs an ordered set of values (voltages, currents, charges, etc.). In other words, there is no phenomena that exhibits more than two independent states. 

In quantum computing, a qubit is the basic unit of quantum information, the counterpart of the binary bit. While the outcome for measurement of a qubit is 0 and 1, the general state of a qubit can be the linear superposition of its two general states. With quantum mechanics, we have multiple-valued units of information, and quantum gates are M-valued circuits. An IBM Quantum computer is even available through the cloud \cite{IBM}. There are very optimistic claims, such as this quote in  \cite{IBM}: ``This is the beginning of the quantum age of computing and the latest advance in IBM towards building a universal quantum computer. A universal computer, once built, will represent one of the greatest milestones in the history of information technology and has the potential to solve certain problems we couldn't solve, and will never be able to solve, with today's classical computers''. At the same time, software limits are noticed \cite{Aaronson}. People think that ``the technology is still in its infancy'' \cite{Tarantola}. 
\subsection{Algorithms}
Several quantum algorithms would solve specific problems, such as factoring integers, exponentially faster than any known classical algorithm. However, there are many others, such as playing chess, proving theorems, scheduling air flights, for which quantum computers would suffer from the same algorithmic limitations as classical computers. They would surpass conventional computers only slightly \cite{Aaronson}. This is why they would be used as coprocessors for the exponentially faster applications.
\subsection{Hardware}
\begin{figure}[htbp]
\centerline{\includegraphics  [width =4 cm]{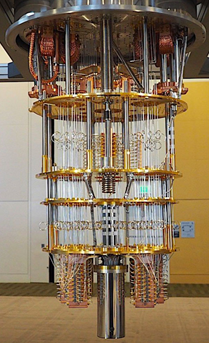}}
\caption{The IBM 50-qubit computer}
\label{QC}
\end{figure}

D-Wave quantum computer has up to 2000 qubits, but the coupling between is limited to restricted local connections (quantum annealing). The 50 qubits of the IBM computer have total interaction. Quantum physics is totally different from classical physics, but operational conditions for quantum devices are also completely different from the conventional integrated circuits. We only list a first features of the IBM quantum computer:
\begin{itemize}
\item The qubits are processed at 15 mK. Different stages operate at 4 K, 800 mK, 100 mK and 15 mK, very close to the absolute zero.
\item The quantum processor is located inside a shield to protect it from electromagnetic radiation
\item The coaxial line between the first and second amplifying stages are made out of superconductors.
\item Quantum amplifiers inside  a magnetic shield capture and amplify processor readout signal while minimizing noise.
\end{itemize}
Fig. \ref{QC} shows that this quantum computer has few similitudes with the computers that we usually use. A previous IBM project with supraconducting technologies in the 80s has shown that it is not easy to make mature a technology close to absolute zero.

Quantum devices are probably the only actual M-valued devices. If quantum computing looks very promising, it will probably remain an important, but small niche (cooled closed to 0\textsuperscript{o} K) when compared to the whole world of personal and IoT devices.

\section {Concluding remarks}
``Advances in integrated circuit technology have been based mostly on CMOS circuit technology operating on the basis of binary logic. However, major problems in present-day LSI technology, such as increased power consumption, interconnection delay limited integration density and device scaling limits, cannot be solved simply  by improving the conventional technology.'' This quote is extracted from \cite{Degawa} published in 2004. Many similar arguments can be found in papers presenting new M-valued circuit designs. Fifteen years later, many improvements have been added to CMOS technologies at technological, circuitry and architectural levels that delay the end of Moore's law and their effects on circuits'performance. And during that period, circuit designers have not considered M-valued circuits as possible solutions to tackle power consumption and interconnection issues for one reason: M-valued circuits can hardly compete when $M>2$!

The main problem for M-valued circuits is that very few devices exhibit a M-valued behavior. With electrical items such as voltages, currents and charges, the set of m values is totally ordered and automatically lead to more complicated circuits. When devices such a single-electron transistor exhibit multiple threshold, this is not sufficient if the corresponding technology is not mature. By now, it seems that quantum qubits are among the very few "multivalued" elements that can be used to build quantum computers that have tremendous performance on some specific applications. However, as long as they will operate at a temperature close to 0\textsuperscript{o} K, these computers will never replace our normal computer environment. With room temperature operation, they would be used as coprocessor for applications on which they outperform classical computers.


\begin{thebibliography}{00}
\bibitem{Gaudet} V. Gaudet, ``A Survey and Tutorial on Contemporary Aspects of Multiple-Valued Logic and Its Application to Microelectronic Circuit'', IEEE Journal on Emerging and Selected Topics in Circuits and Systems, Vol. 6. N\textsuperscript{o} 1. March 2016.
 \bibitem{Gulak} G. Gulak, ``A review of multiple-valued memory technology'', Proceedings. 1998 28th IEEE International Symposium on Multiple- Valued Logic, Fukuoka, 1998, pp. 222-231.
\bibitem{b1} D. Etiemble and M. Israel, ``Comparison of binary and multivalued integrated circuits according to VLSI criteria'', IEEE Computer, pp.28-42, April 1988.
\bibitem{b2} D. Etiemble, Why M-Valued Circuits are restricted to a Small Niche'', in Journal of Multiple Valued Logic and Soft Computing, Vol. 9, N\textsuperscript{o}1, 2003.
\bibitem{Hurst} S.L. Hurst, ``Multiple-Valued Logic - Its Status and Its Future``, IEEE Trans. on Computers, VOL. C-33, N\textsuperscript {o} 12, December 1984.
\bibitem{b3} F. Pollack, ``New Microarchitecture Challenges in the Coming Generations of CMOS Process Technologies'', Intel Corp. Micro32 conference key note - 1999.
\bibitem{b4} D. Etiemble, and M. Israel, ``A new concept for ternary logic elements'', Proc. 1974 International  Symposium on Multiple Valued  Logic. Morgantown, West Virginia, pp 437-456.
\bibitem{b5} M. Brilman, D. Etiemble, J.L.Oursel , and P.Tatareau, ``A 4-valued ECL Encoder and Decoder Circuit'', IEEE J. Solid State Circuits, Vol. SC17, N\textsuperscript{o} 3, pp 547-552, June 1982.
\bibitem{b6}T. Dao, L. Russell, D. Preedy, and E. McCluskey, ``Multilevel I2L with threshold gates,'' 1977 IEEE International Solid-State Circuits Conference. Digest of Technical Papers, Philadelphia, PA, USA, 1977, pp. 110-111.

\bibitem{Aoki} M. Aoki et al, ``A 16-level/Cell Dynamic Memory'', IEEE J. Solid State Circuits, Vol. SC-22, N\textsuperscript{o} 2, pp. 297-299, April 1987.
\bibitem{Sugi} T. Sugibayashi, I. Naritake, A. Utsugi et al, ``A 1 Gb DRAM for File Application'', ISSCC Digest of Technical Papers, pp. 254-255, Feb. 1995.
\bibitem{Murotani} T. Murotani et al., ``A 4-level Storage 4Gb DRAM'', ISSCC Digest of Technical Papers, pp.74-75, Feb. 1997.

\bibitem{Bauer} J. Bauer et al, ``A multilevel-Cell 32Mb Flash Memory'', ISSCC Digest of Technical Papers, pp.132-133, Feb. 1995.
\bibitem{Jung} T. Jung et al, ''A 3.3 V 128 Mb Multi-Level Flash Memory for Mass Storage Applications'', ISSCC Digest of Technical Papers, pp.32-33, Feb. 1996.
\bibitem{Shulaker} M.M. Shulaker, G. Hills, N. Patil, H. Wei, H.Y Chen, H.S. Philip Wong and S. Mitra, ``Carbon nanotube computer", Nature, Vol 501, 26 September 2013, doi: 10.1038/nature12502.
\bibitem{Navi} F. Sharifi, et al., ``Robust and energy-efficient carbon nanotube FET-based MVL gates: A novel design approach'', Microaelectronic. J (2015), http://dx.doi.org/10;1016/j.mejo.2015.09;018.
\bibitem{CEA} V.  Deshpande. ``Scaling Beyond Moore: Single Electron Transistor and Single Atom Transistor
Integration on CMOS.'' Ph.D Universit\'e de Grenoble, 2012, https://tel.archives-ouvertes.fr/tel-00813508/document.
\bibitem {Degawa} K. Degawa, T. Aoki, T.Higuhi, H. Inokawa and Y. Takahashi,  ``A Single Electron Transistor Logic Gate Family for Binary, Multivalued and Mixed-Mode Logic'', IEICE Trans. Electron., Vol E87-C, N\textsuperscript{o} 11, November 2004
\bibitem{Batt} D. Bhattacharjee, W. Kim, A. Chattopadhyay, R. Waser and V. Rana, ``Multi-valued and Fuzzy Logic Realization using TaOx Memristive Devices", Scientific Reports 8, Article number: 8 (2018), DOI https://doi.org/10.1038/s41598-017-18329-3
\bibitem{IBM} ``IBM Makes Quantum Computing Available on IBM Cloud to Accelerate Innovation'', IBM News room, 2016-05-04, https://www-03.ibm.com/press/us/en/pressrelease/49661.wss.
\bibitem{Aaronson} S. Aaronson, ``The limits fo Quantum'', Scientific American, pp. 62-69, March 2008.
\bibitem {Tarantola} A. Tarantola, ``Not even IBM is sure where its quantum computer experiments will lead''. Personal Computing, https://www.engadget.com/2018/02/23/ibm-q-quantum-computer-experiments.



\end{thebibliography}
\end{document}